\documentclass[twocolumn]{aastex631}  

\usepackage[capitalise]{cleveref}
\usepackage{comment}
\usepackage{graphicx}
\usepackage{appendix}
\usepackage{multirow}
\usepackage[]{footmisc}
\usepackage[caption=false]{subfig}	

\shorttitle{Theoretical modeling of GRB~221009A}
\shortauthors{L.Foffano, M.Tavani, G. Piano}

\begin{document}

\title{Theoretical modeling of the exceptional GRB~221009A afterglow}

\correspondingauthor{Luca Foffano}
\email{luca.foffano@inaf.it}

\author[0000-0002-0709-9707]{Luca Foffano}
\affiliation{INAF-IAPS Roma, via del Fosso del Cavaliere 100, I-00133 Roma, Italy}

\author[0000-0003-2893-1459]{Marco Tavani}
\affiliation{INAF-IAPS Roma, via del Fosso del Cavaliere 100, I-00133 Roma, Italy}
\affiliation{Dip. di Fisica, Università di Roma Tor Vergata, via della Ricerca Scientifica 1, I-00133 Roma, Italy}

\author[0000-0002-9332-5319]{Giovanni Piano}
\affiliation{INAF-IAPS Roma, via del Fosso del Cavaliere 100, I-00133 Roma, Italy}

\received{July 24$^{\text{th}}$, 2024}
\revised{August 23$^{\text{rd}}$, 2024}
\accepted{August 31$^{\text{st}}$, 2024}

\begin{abstract}
The extraordinary gamma-ray burst GRB~221009A provides a great opportunity to investigate the enigmatic origin and evolution of GRBs. However, the complexity of the observations associated with this GRB provides significant challenges to developing a theoretical modeling in a coherent framework.
In this paper, we present a theoretical interpretation of the GRB~221009A afterglow within the relativistic fireball scenario, aiming to describe the broadband dataset with a consistent model evolution. We find that the adiabatic fireball evolution in the slow-cooling regime provides a viable scenario in good agreement with observations.
Crucial to our analysis is the set of simultaneous GeV and TeV gamma-ray data obtained by AGILE and LHAASO during the early afterglow phases.
Having successfully modeled as inverse Compton emission the high-energy spectral and lightcurve properties of the afterglow up to $10^4$~s, we extend our model to later times when also optical and \mbox{X-ray} data are available. This approach results in a coherent physical framework that successfully describes all observed properties of the afterglow up to very late times, approximately $10^6$~s. Our model requires time-variable microphysical parameters, with a moderately increasing efficiency $\varepsilon_e$ of a few percent for transferring the shock energy to radiating particles and a decreasing efficiency for magnetic field generation $\varepsilon_B$ in the range $10^{-5}$-$10^{-7}$. Fitting the detailed multifrequency spectral data across the afterglow provides a unique test of our model.
\end{abstract}



\section{Introduction} 
\label{sec:intro}
\noindent
GRB~221009A is the brightest gamma-ray burst (GRB) ever detected. Its origin is connected with the core collapse of a massive star 
\citep{Srinivasaragavan2023, supernova_grb221009a} at redshift $z = 0.15095 \pm 0.00005$ \citep{2023arXiv230207891M, 2022GCN.32648....1D}. With an unprecedented brightness and duration, GRB~221009A offers an exceptional opportunity to investigate the physical mechanisms driving such powerful explosions. 

On 2022 October $9$, the transient event \textit{Swift}~J1913.1+1946 \citep{2022ATel15651....1N}  was then identified as GRB~221009A \citep{2022GCN.32636....1V} and associated with the reference \textit{\textit{Fermi}}-GBM trigger time $T_{0,\text{GBM}}$=13:16:59.99~UT \citep{lesage_2022}. 
Several instruments captured the event during its prompt emission, including \textit{Konus}-Wind \citep{Frederiks_2023} and \textit{Fermi}-GBM \citep{fermi_gbm_grb221009a}. The afterglow was systematically observed and monitored in the following days \citep{2022GCN.32652....1B, 2022GCN.32662....1K, 2022ATel15651....1N}.

\begin{figure*}[t]
    \centering
    \includegraphics[width=0.8\textwidth]{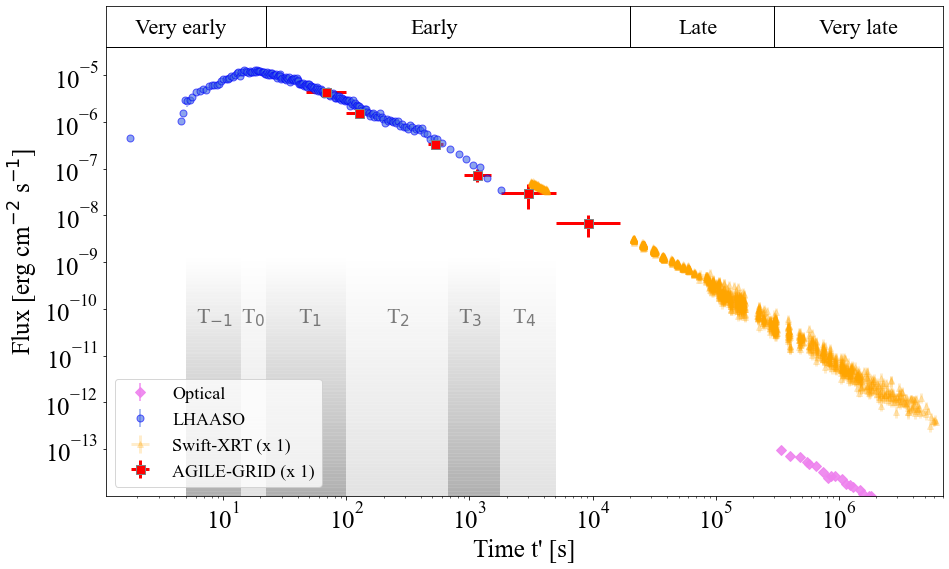}
    \caption{Flux intensity lightcurves of the GRB~221009A afterglow for different energy bands. We include  optical data (pink diamonds), X-ray data (orange triangles), GeV data (red squares), and TeV data (blue circles). We indicate the relevant time intervals considered in our analysis as given in \Cref{tab:evolution_parameters}. We also define four main phases of the afterglow (very early, early, late, and very late).}
        \label{fig:lightcurve_only_data}
\end{figure*}

The AGILE satellite \citep{tavani_2009_agile_mission} detected GRB~221009A during its most important phases, providing valuable information in the MeV and GeV energy range with the MCAL and GRID instruments \citep{Tavani_2023}.  TeV gamma rays were also reported during the initial phases of GRB~221009A by the LHAASO observatory \citep{LHAASO_2023}. The firm simultaneous detection at MeV, GeV and TeV gamma rays - together with data at other energies - provides crucial information for the modeling of this powerful event.
 
In this paper, we focus on the GRB~221009A afterglow, omitting a detailed study of the early-phase prompt emission, which is beyond the scope of our investigation.
In Section~\ref{sec:mwl_data} we briefly summarize the multi-wavelength datasets  adopted in this analysis and focus on the simultaneous data collected by AGILE and LHAASO.
In Section~\ref{sec:modeling}, we describe the details of the theoretical model. Then, Sections~\ref{sec:spectra} and \ref{sec:lightcurve} are devoted to present the comparison between the model and the multiband spectral and flux intensity data. Finally, in Section~\ref{sec:discussion} we briefly discuss the physical results of our analysis. 

\section{Multi-wavelength data}
\label{sec:mwl_data}
\noindent
Our goal is to properly consider all relevant information regarding the flux and spectral evolution  of the multiband datasets of the GRB~221009A afterglow. In \Cref{fig:lightcurve_only_data} we show the lightcurves
from different instruments and energy bands: TeV gamma-ray data from the LHAASO observatory \citep{LHAASO_2023}, GeV gamma-ray data by AGILE-GRID (described in the next section), X-ray data provided by \textit{Swift}-XRT \citep{Swift_williams_2023}, and optical data\footnote{In this analysis, we do not report the \textit{Swift}-UVOT data, as their flux is in contrast with the collimation-corrected optical data. We attribute this effect to a strong Galactic absorption, typical for this energy band.} obtained with Pan-STARRS and other imaging facilities by \citet{Fulton_optical_lightcurve}. 

\begin{figure*}[t]
    \centering
    \subfloat[Interval $T_{-1}$.]{
    \includegraphics[width=1.06\columnwidth]{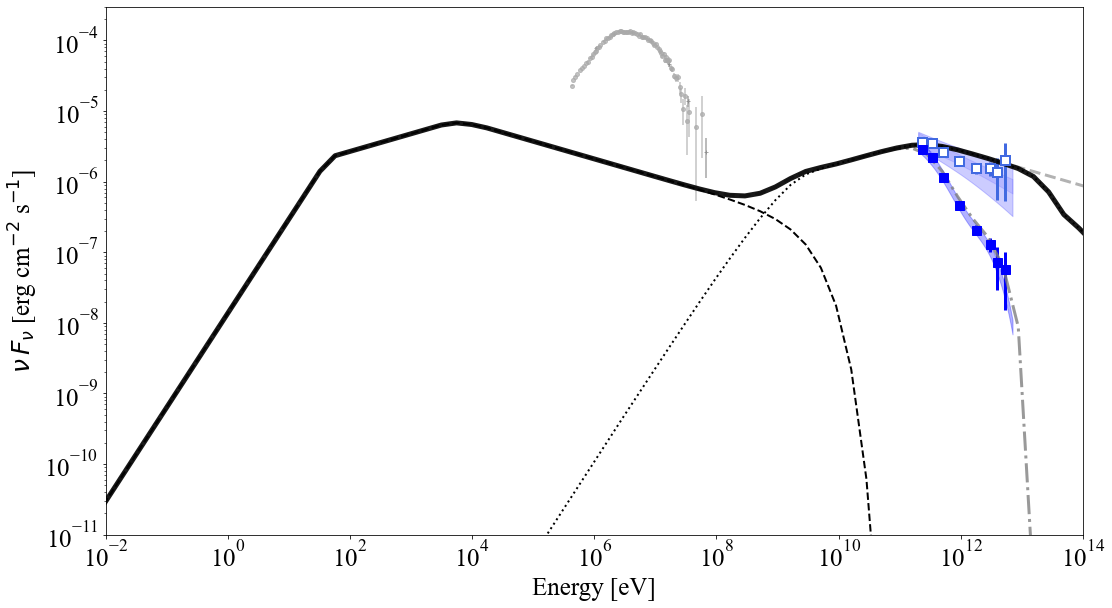}
    \label{fig:sed_AGILE_LHAASO_int-1}}
    \subfloat[Interval $T_0$.]{\includegraphics[width=1.06\columnwidth]{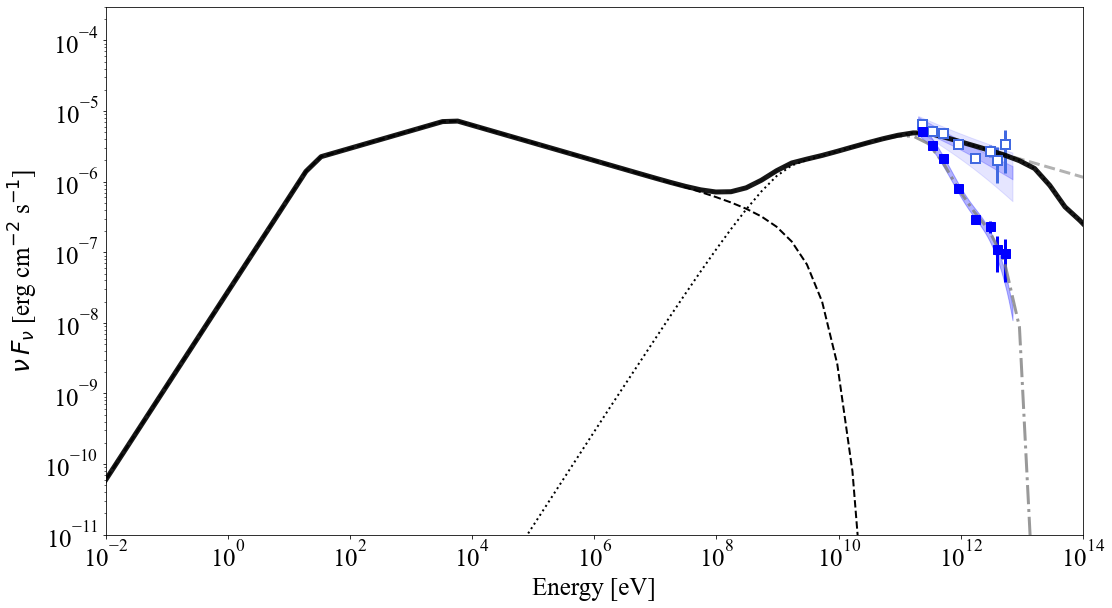}\label{fig:sed_AGILE_LHAASO_int0}
    }
    \caption{Spectral energy distributions calculated for the time intervals $T_{-1}$ and $T_0$, with parameters reported in \Cref{tab:evolution_parameters}. The overall SSC emission (continuous thick line) is given by synchrotron (black dashed) and inverse Compton (black dotted) contributions. The TeV SSC emission without $\gamma\gamma$ internal absorption is shown with a dashed and the EBL absorbed emission is shown with a dashed–dotted gray line, and the EBL absorbed emission is shown with a dashed–dotted gray line. LHAASO data points are reported in blue squares, both the observed data (dark) and the EBL deabsorbed data (empty). The corresponding butterflies are related to the statistical error (dark) and to the additional error due to the EBL de-absorption process (faded). AGILE MCAL data at $t' = [-15:-3]$ s are shown in gray in (a).} 
    \label{fig:sed_AGILE_LHAASO_very_early_times}
\end{figure*}

In the late-time spectra we include the hard X-ray data provided by the BAT instrument on board the \textit{Swift} telescope \citep{Swift_williams_2023}. 
Concerning the TeV LHAASO spectral data\footnote{We do not show the spectra up to 10 TeV obtained by LHAASO in \citet{LHAASO_data_above_10tev}.
 However, they substantially support our scenario.} shown in Figures~\ref{fig:sed_AGILE_LHAASO_very_early_times} and \ref{fig:sed_AGILE_LHAASO_early_times}, we show the observed data and the data deabsorbed for the extragalactic background light (EBL).

Since a minor X-ray precursor of GRB 221009A was detected at the nominal trigger time  $T_{0, \text{GBM}}$ (which turned out to occur at a much earlier time than that of the main GRB episode), it is convenient  to  define the time $T^* = T_{0, \text{GBM}}$ + 226~s and adopt in our analysis the renormalized time 
$t' = t - T^*$  \citep{lazzati_2006_t0_afterglow, Kobayashi_2007_t0_afterglow, LHAASO_2023, Khangulyan2024}.
From now on, we use $t'$, if not differently indicated.

In \Cref{fig:lightcurve_only_data} we define the different afterglow phases and the relevant time intervals 
$T_{-1}, T_{0}, T_1, T_2, T_3, T_4$ as defined in \Cref{tab:evolution_parameters}.
In our calculations, we assume cosmological parameters describing a flat Universe
 with $\Omega_{M}$ = 0.3, $\Omega_{\Lambda}$ =
0.7 and $H_{0}$ = 70~km s$^{-1}$~Mpc$^{-1}$.


\subsection{AGILE data}
\label{sec:AGILE_data_analysis}
\noindent
We performed a specific analysis of AGILE-GRID GeV data strictly simultaneous with LHAASO TeV data. The procedure is analogous to that presented in \citet{Tavani_2023}, and was performed by dividing the AGILE observations into time intervals consistent with those reported in \citet{LHAASO_2023}. 
The data analysis takes into account only the effective GRID exposure, given the fact that the instrument was exposed to the GRB discontinuously due to the AGILE telescope's spinning.
The GRID spectra - shown in Figures \ref{fig:sed_AGILE_LHAASO_int1}, \ref{fig:sed_AGILE_LHAASO_int2}, and \ref{fig:sed_AGILE_LHAASO_int3} - were obtained from 50 MeV up to the maximum energies allowed by  photon statistics.
The lightcurve - shown in Figures~\ref{fig:lightcurve_only_data} and \ref{fig:lightcurve} - was generated in the energy range 50~MeV - 3~GeV. In each time bin, the energy flux was computed by assuming the specific power-law photon index of the interval, whenever the photon statistics allowed a proper spectral analysis (otherwise adopting a photon index of 2). All details are reported in \Cref{tab:table_observ_windows}.\\


\newcommand\setrow[1]{\gdef\rowmac{#1}#1\ignorespaces}
\newcommand\clearrow{\global\let\rowmac\relax}
\clearrow
\begin{table}
\centering
\caption{Time-dependent quantities adopted in 
the relativistic fireball modeling of GRB~221009A. The global parameters used in the model are $E_{\text{iso,0}} = 7\cdot10^{55}$ erg, $\Gamma_0 = 480$, s = 0, $\text{n}_0=0.8$ cm$^{-3}$, p $= 2.5$, and $\gamma_{\text{max}} = 4 \cdot 10^7 $.}
\renewcommand{\arraystretch}{1.3}
\begin{tabular}{|ccccc|}
\toprule
\multicolumn{5}{|c|}{\textbf{Time-dependent Quantities}}\\
\setrow{\bfseries}Interval & $t'_{start}$ [s]  & $t'_{stop}$ [s]  &  $\varepsilon_e(t')$ & $\varepsilon_B(t')$  \\ 
\hline
$T_{-1}$ & 5 & 14 &  1.7$\cdot 10^{-2}$ & 2.0$\cdot 10^{-5}$ \\ 
$T_0$ & 14 & 22 &     1.9$\cdot 10^{-2}$ & 1.1$\cdot 10^{-5}$  \\ 
$T_1$  & 22 & 100 &   2.0$\cdot 10^{-2}$ & 6.0$\cdot 10^{-6}$  \\ 
$T_2$  & 100 & 674   &   2.9$\cdot 10^{-2}$ & 1.0$\cdot 10^{-6}$  \\ 
$T_3$  & 674 & 1774  &   3.8$\cdot 10^{-2}$ & 3.0$\cdot 10^{-7}$   \\ 
$T_4$  & 1788 & 5043 &   5.0$\cdot 10^{-2}$ & 1.4$\cdot 10^{-7}$   \\
\toprule
\end{tabular}
\label{tab:evolution_parameters}
\end{table}

\section{The relativistic fireball model}
\label{sec:modeling}
\noindent
The relativistic fireball model provides a theoretical framework to study the afterglow emission of a blast wave expanding in an external environment \citep[e.g.,][]{Rees_1992_relativistic_fireball, Piran_review_GRB_1998, Chiang_dermer_1999, Panaitescu_2000A}. This interaction develops a forward shock propagating outwardly and a reverse shock propagating into the shell.
The forward shock front expands spherically with an initial bulk Lorentz factor $\Gamma_0$, following a hydrodynamic evolution $\Gamma(t)$ with analytic solutions in the fully adiabatic or radiative scenarios of the expansion \citep{Blandford_1976, sari_hydrodynamics}. The relation between time and radial distance $r$ in the observer's frame is  given by $r = 4 \Gamma^2 c t$ \citep{sari_1998}.

The shock front propagates into the surrounding interstellar medium (ISM)  with a density profile in the rest frame $n(r) = n_0 \, r^{-s}$, with $s = 0$  for a homogeneous environment and $s = 2$ 
 for a surrounding medium determined by a massive star wind density profile.
In the fully adiabatic hydrodynamic evolution, these two scenarios cause the Lorentz factor to change over the radial distance (and time) as $\Gamma(r) \propto r^{-3/2}$ in the homogeneous scenario and as $\Gamma(r) \propto r^{-1/2}$ in the wind-like case.

As the blast wave decelerates, fractions of the shock energy are transferred on a short timescale to the magnetic field and to the random kinetic electron energy through the quantities $\varepsilon_B$ and $\varepsilon_e$, respectively. Given their importance in our analysis, it is worth briefly reminding the reader of the physical definitions of $\varepsilon_B$ and $\varepsilon_e$ that are ultimately determined by the energy density $U_{\text{sh}}$ in the forward shock. The initial kinetic energy of the blast wave $E_{k}$ is carried mostly by the protons. 
The shock energy density in the bulk frame can be represented as $U_{\text{sh}} = 4\Gamma^2 \, n_{p1} \, m_p \, c^2$, where $n_{p1}$ is the rest-frame upstream proton number density and $m_p$ the proton mass, and we have adopted a shock compression ratio of $4 \Gamma$. The magnetic field $B$ in the comoving forward shock frame is obtained from the relation $U_B = \varepsilon_B \, U_{\text{sh}}$, with $U_B = {B^2}/{8\pi}$ the magnetic field energy density. 
We therefore have, 
\begin{equation}
      4\Gamma^2 \, n_{p1} \, m_p \, c^2 \, \varepsilon_B = \frac{B^2}{8 \, \pi} \;,
\end{equation}
and then $B =  \Gamma c \, \sqrt{32 \, \pi \, n_{p1} \, m_p \, \varepsilon_B}$.

Electrons and positrons absorb a fraction $\varepsilon_e$ of the comoving energy density $U_{\text{sh}}$, obtaining the electron energy density $U_e = U_{\text{sh}} \, \varepsilon_e$. 
They are accelerated, and a power-law energy distribution $dN (\gamma) / d\gamma = \kappa \, \gamma^{-p}$ is established on a timescale shorter than the dynamical timescale, with $p$ the power-law index and $\kappa$ the normalization factor. The electrons' Lorentz factor $\gamma$ ranges from a minimum value $\gamma_{\text{min}}$ to a maximum value $\gamma_{\text{max}}$. Their energy density becomes $U_e = \kappa \, m_e \, c^2 \int_{\gamma_{\text{min}}}^{\gamma_{\text{max}}} \gamma \, \frac{\text{d}N}{\text{d}\gamma} \, \text{d}\gamma$.\\
Assuming that $\gamma_{\text{max}} \gg \gamma_{\text{min}}$, we get:
\begin{equation}
    \Gamma \, n_{p2} \, m_p \, c^2 \, \varepsilon_e = n_{e2} \, m_e \, c^2 \, \frac{(p-1)}{(p-2)} \gamma_{\text{min}} \: ,
    \end{equation}
where $n_{p2} = 4 \Gamma n_{p1}$ and $n_{e2}$ are the downstream proton and electron number densities in the rest frame.
It is commonly assumed that $n_{p2} = n_{e2}$ and that the complexity of the particle acceleration process gets absorbed into the quantity $\varepsilon_e$, leading to the relation $\gamma_{\text{min}} = \frac{p - 2}{p - 1} \, \frac{m_p}{m_e} \, \varepsilon_e \, \Gamma$ \citep{sari_cooling_timescales}.
We anticipate that, unlike in most GRB models, in the case of GRB~221009A both quantities $\varepsilon_B$ and $\varepsilon_e$ are required to vary with time as a consequence of the fundamental physical processes determining the particle energy evolution of a very complex and long event.

As the fireball expands, the accelerated electrons and positrons radiate their energy by synchrotron and inverse Compton emission through the Synchrotron Self-Compton process \citep[SSC, e.g.,][]{sari_1998, sari_esin_2001}. 
The relation between $\gamma_{\text{min}}$ and the cooling Lorentz factor $\gamma_c = \frac{6 \pi m_e c}{\sigma_T \Gamma B^2 t}$ is crucial, and defines two distinct physical regimes. When $\gamma_{\text{min}} > \gamma_c$, particles are in a \textit{fast-cooling}  regime, efficiently losing their energy through synchrotron cooling within a dynamical time. Conversely, when $\gamma_{\text{min}} < \gamma_c$, particles are in a \textit{slow-cooling} regime, and only particles with $\gamma > \gamma_c$ cool efficiently.

In our model, we account for internal $\gamma\gamma$ absorption and cosmological effects. We also include corrections for Klein-Nishina scattering \citep{Nakar2009} though these are mostly negligible for the modeling of this event. Additionally, we consider the absorption effects due to interactions with the EBL, adopting the model by \citet{ebl_dominguez_2011}.
We include the cooling effect due to the inverse Compton process, which shortens the electrons' cooling time. The previously defined cooling Lorentz factor - now named $\gamma_{\text{c,syn}}$ - is then modified as $\gamma_c = \gamma_{\text{c,syn}} / (1+Y)$, where $Y$ is the Compton parameter computed following \citet{sari_esin_2001}.

\subsection{Study cases}   
\noindent
In our study, we have investigated several alternative scenarios. We considered both a radiative and an adiabatic evolution, the latter both within a fast- and a slow-cooling regime.
Applications of these scenarios to the precise modeling of the multiband spectral and intensity data of GRB~221009A turned out to be a quite challenging task, with model-dependent outcomes that often resulted in contradiction with the data.

In this paper, we restrict our analysis to a successful scenario that explains both the GeV-TeV spectral and intensity evolution of the early afterglow as well as the X-ray and optical evolution of the late afterglow. We find that a fully adiabatic evolution of the fireball in a homogeneous medium with nonconstant values of $\varepsilon_e$ and $\varepsilon_B$ moderately changing with time is remarkably successful in explaining the overall afterglow within a single evolution scenario. \\

\begin{figure*}[t]
    \centering
    \subfloat[Interval $T_1$.]{
    \includegraphics[width=1.05\columnwidth]{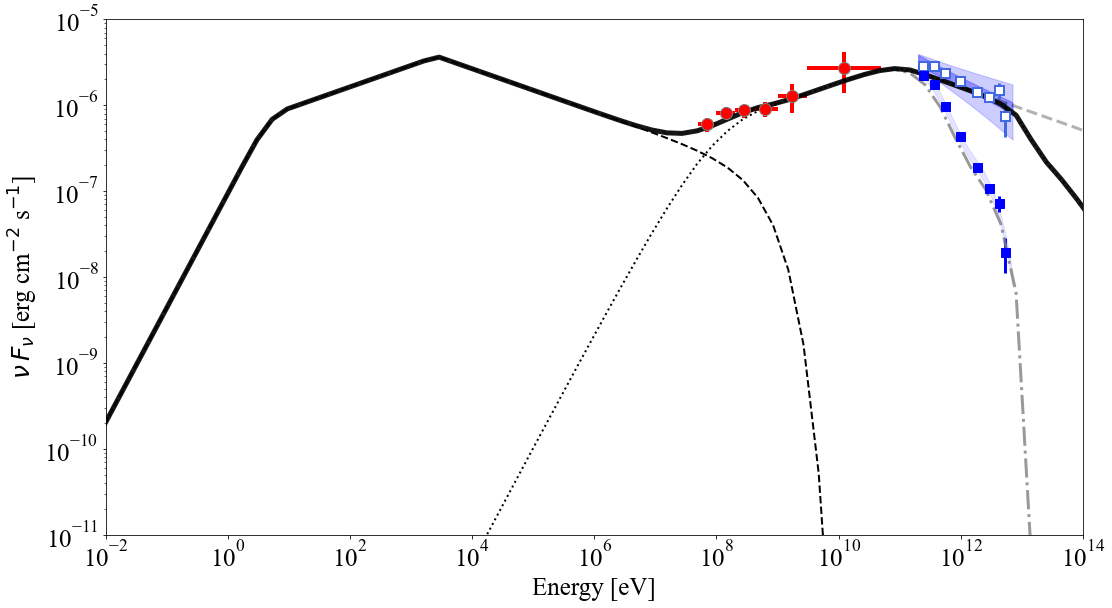}
    \label{fig:sed_AGILE_LHAASO_int1}}
    \subfloat[Interval $T_2$.]{\includegraphics[width=1.05\columnwidth]{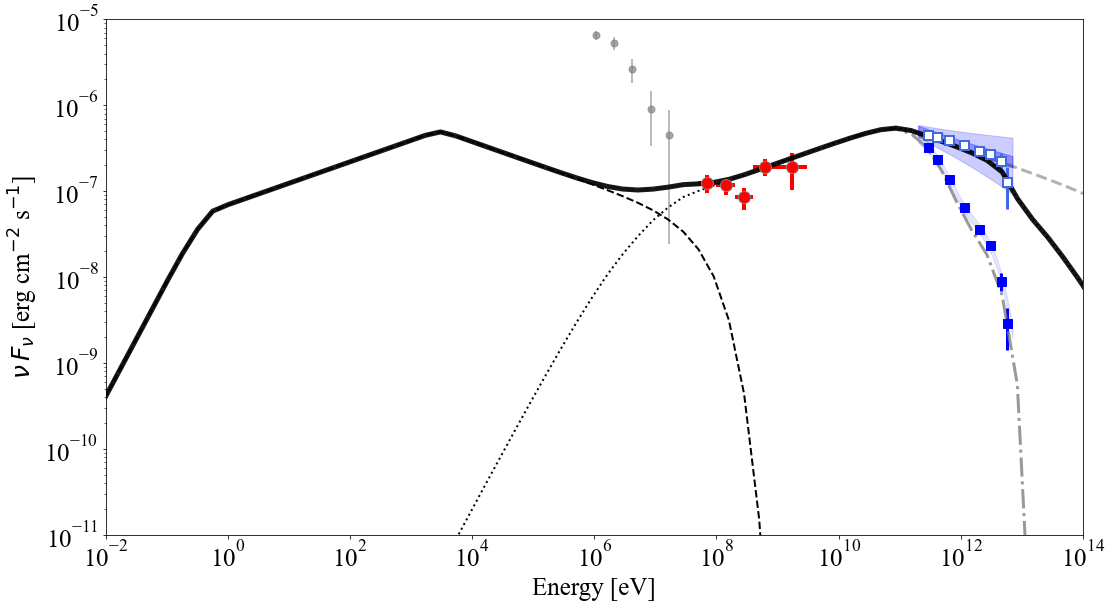}\label{fig:sed_AGILE_LHAASO_int2}}\\
    \subfloat[Interval $T_3$.]{\includegraphics[width=1.05\columnwidth]{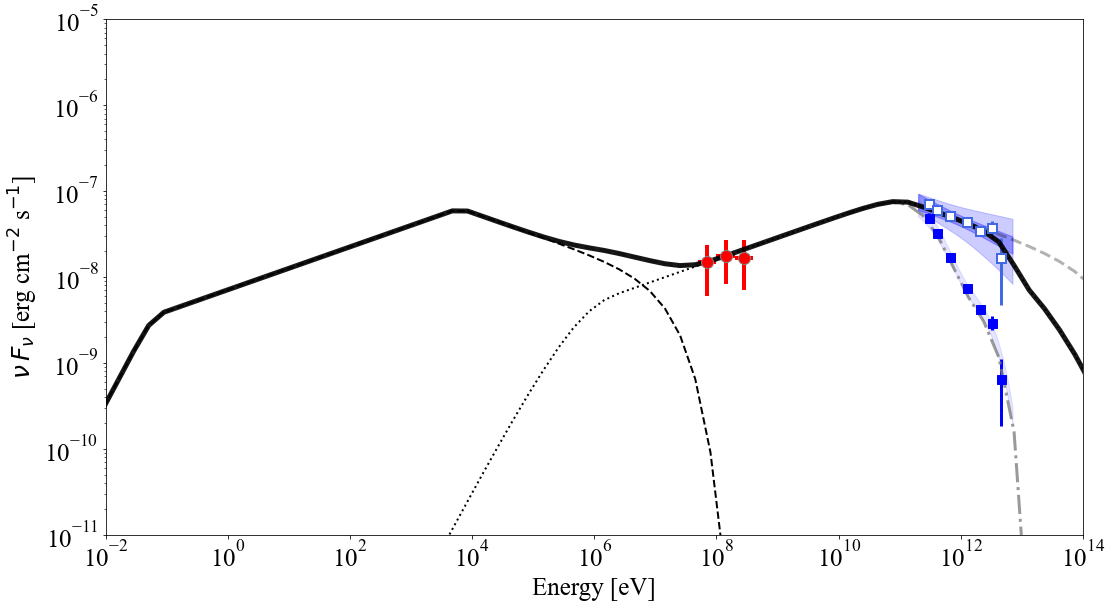}\label{fig:sed_AGILE_LHAASO_int3}}
    \subfloat[Interval $T_4$.]{\includegraphics[width=1.05\columnwidth]{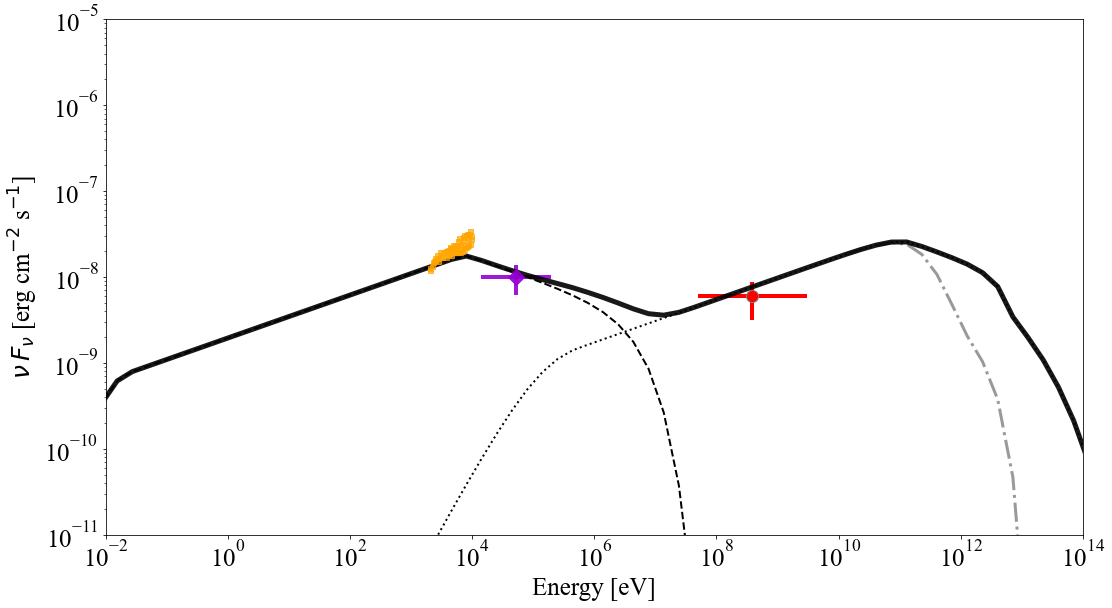}\label{fig:sed_AGILE_LHAASO_int4}}
    \caption{Calculated spectral energy distributions (as in \Cref{fig:sed_AGILE_LHAASO_very_early_times}) for the time intervals $T_1$, $T_2$, $T_3$, and $T_4$. We show simultaneous data from AGILE GRID (red circles) and LHAASO (blue squares). AGILE MCAL data at $t' = [167:256]$ s are shown in gray in (b). In the $T_4$ interval, also X-ray and hard X-ray data from \textit{Swift}-XRT (orange squares) in $t' = [3174:4274]$~s and BAT (violet diamond) in $t' = [3674:4301]$~s  are reported.}   
    \label{fig:sed_AGILE_LHAASO_early_times}
\end{figure*}


\section{Spectral analysis}
\label{sec:spectra}

\noindent
Matching with good accuracy the results of our modeling for the very early and early afterglow phases with the multiband spectra and intensity evolution was our first goal. This task required a global fitting of the available datasets, mainly driven by the unique set of simultaneous spectral data in the GeV-TeV ranges as provided by AGILE and LHAASO. 

Figures~\ref{fig:sed_AGILE_LHAASO_very_early_times} and \ref{fig:sed_AGILE_LHAASO_early_times} show the entire set of available spectral data and the best theoretical modeling. \Cref{tab:evolution_parameters} provides the time-dependent quantities of the modeling for the relevant time intervals.  The adopted global parameters are $E_{\text{iso,0}} = 7\cdot10^{55}$ erg, $\Gamma_0 = 480$, s = 0, $\text{n}_0=0.8$ cm$^{-3}$, p $= 2.5$, and $\gamma_{\text{max}} = 4 \cdot 10^7 $. 
The optimized set of parameters for our model, including their time evolution, was obtained by a {\it global} analysis of the very early and early phases lasting up to about $t' = 10^4$~s. We also extended our analysis to late times, as discussed below.

\subsection{The very early afterglow phase}
\label{sec:MCALdata}
\noindent
In the first afterglow phase, the earliest spectral data available for this GRB are given by the two LHAASO spectra within $t' = [5:14]$~s and $[14:22]$~s (no AGILE GeV data are available for these intervals because of exposure and saturation).

In Figures~\ref{fig:sed_AGILE_LHAASO_int-1} and \ref{fig:sed_AGILE_LHAASO_int0}, we show the spectral energy distributions (SEDs) of the model, which are in agreement with both the unabsorbed and the EBL-absorbed LHAASO data. Considering the model parameters adopted for these two intervals reported in \Cref{tab:evolution_parameters}, it is interesting to note that a successful spectral modeling requires $\varepsilon_B \ll \varepsilon_e$ and $\gamma_{\text{min}} \ll \gamma_c$ since the beginning of the afterglow.

Furthermore, in \Cref{fig:sed_AGILE_LHAASO_int-1}, we add the AGILE MCAL data between $t' = [-15:-3]$~s, just before the main burst after which the MCAL instrument was saturated. Although they are not   simultaneous with the GeV-TeV data, they confirm that the prompt MeV emission was significantly more intense than the afterglow synchrotron emission predicted by the model. Indeed, we interpret the MeV emission as an additional component related with the prompt phase of the GRB, and not to the afterglow. Spectra obtained a few seconds later by the Konus-Wind instrument in \citet{Frederiks_2023} between $t' = [-1:7]$ s support this interpretation. As indicated by \textit{Fermi-}GBM \citep{lesage_2022} and Konus-Wind data, the hard X-ray/MeV emission progressively decreased with time, allowing the underlying afterglow  emission to emerge in this energy range after several hundreds of seconds.

\begin{figure*}[t]
    \centering
    \includegraphics[width=0.8\textwidth]{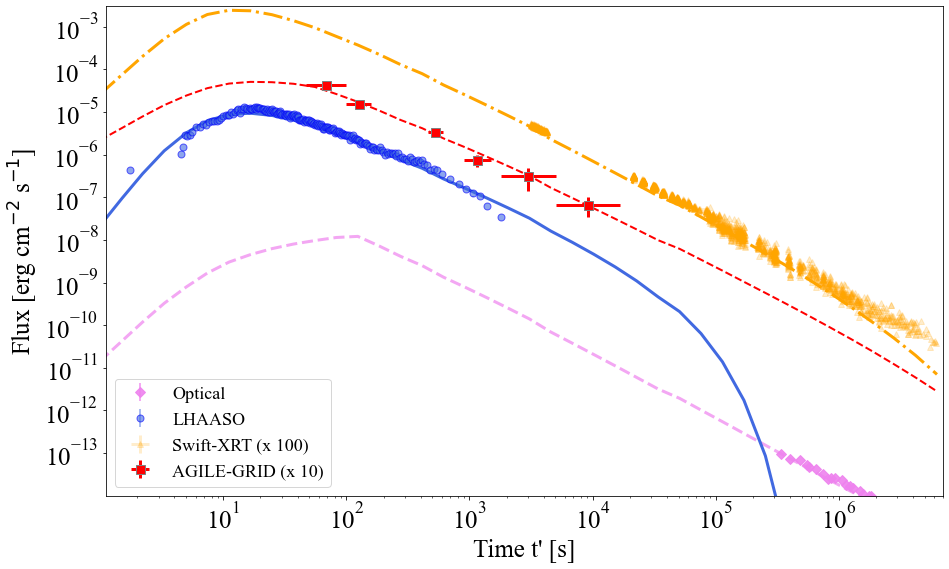}
    \caption{Calculated lightcurves for different energy bands. As in \Cref{fig:lightcurve_only_data}, we include  optical data (pink diamonds), \textit{Swift}-XRT X-ray data (orange triangles, multiplied by $100$), AGILE-GRID GeV data (red squares, multiplied by $10$), and LHAASO TeV data (blue circles). The TeV calculated lightcurve is shown after the EBL de-absorption.}
        \label{fig:lightcurve}
\end{figure*}

\subsection{The early afterglow phase: simultaneous GeV-TeV data}
\label{sec:datasets}
\noindent
GeV and TeV gamma-ray spectral data during the early afterglow ($T_1$, $T_2$, and $T_3$ intervals) between $t' = [20: 10^4]$~s turn out to be crucial for our analysis.
Figures~\ref{fig:sed_AGILE_LHAASO_int1}-\ref{fig:sed_AGILE_LHAASO_int3} show such datasets and our optimized spectral modeling during this phase.

In our interpretation, the quantities $\varepsilon_e$ and $\varepsilon_B$ are required to change in time for a precise fitting between model and spectral data. In intervals $T_1-T_3$, the quantity $\varepsilon_e$ slightly increases from $2.0\cdot 10^{-2}$ to $3.8 \cdot 10^{-2}$, and the quantity $\varepsilon_B$ diminishes from $6\cdot 10^{-6}$ to
$3\cdot 10^{-7}$. This time dependence reflects the evolution of the physical conditions determining the particle acceleration and magnetic field efficiencies. Keeping the quantities $\varepsilon_e$ and $\varepsilon_B$ constant throughout the early afterglow phase leads to a significant discrepancy between model and data. The model with parameters given in \Cref{tab:evolution_parameters} and Section~\ref{sec:spectra} provides a viable framework: the agreement between data and model is satisfactory despite the stringent spectral constraints. As we will see, this approach is successful also in the subsequent phases of the afterglow.

The GeV-TeV component is interpreted as inverse Compton emission. Very interestingly, during these GeV-TeV simultaneous observations lasting from about 20 to more than $600$~s, the position of the inverse Compton peak is almost constant in time. This represents an important constraint for the modeling, as we will discuss below.

In the $T_2$ interval in \Cref{fig:sed_AGILE_LHAASO_int2}, we also show simultaneous AGILE MCAL spectral data. As mentioned earlier, we confirm that they belong to the decaying prompt component and that no information on the synchrotron afterglow component in this energy range can be obtained at this time. 

\subsection{The early afterglow phase: simultaneous X-ray and GeV data}
\noindent
New important spectral data become available at $t' \geq 3000$~s during the $T_4$ interval - shown in \Cref{fig:sed_AGILE_LHAASO_int4} - in the X-ray (in $t' = [3174:4274]$~s) and hard X-ray (in $t' = [3674:4301]$~s) ranges.
These data are the first set of observations of the \textit{Swift} telescope, and are simultaneous with the AGILE observations. For this reason, they represent an important opportunity to constrain - for the first time in the GRB~221009A afterglow - the crucial synchrotron peak.
The \textit{Swift}-XRT instrument reported a hard photon index of $1.61 \pm 0.02$, while the \textit{Swift}-BAT instrument reported a softer photon index of $2.13 \pm 0.19$. This indicates a spectral break between 10 and 100 keV, which then constrains the peak of the synchrotron emission and consequently the model parameters and their evolution.
Furthermore, the X-ray and hard X-ray emissions can provide useful information on the relation between the overall synchrotron component versus the inverse Compton emission. 
During the $T_4$ interval, no strictly simultaneous TeV data are available. However, the very last TeV point in the LHAASO lightcurve corresponds to the beginning of this time interval and substantially confirms our modeling at these times.

\section{Lightcurve modeling}
\label{sec:lightcurve}
\noindent
\Cref{fig:lightcurve} presents the results of our model with the computation of the GRB~221009A afterglow lightcurves in different energy intervals spanning over the optical, \mbox{X-ray}, GeV, and TeV bands. The observed lightcurves, described in Section~\ref{sec:mwl_data}, are also shown. 
Overall, a remarkable agreement between our fireball model and the data is verified over a very long time interval.

At early times $t' < 10^4$~s, both the GeV and TeV gamma-ray data evolution are well matched by our model. 
Notably, no jet breaks can be identified in either our model of the TeV emission (which continues with the same time evolution up to $t' \sim 10^4$~s and then decays) or in the AGILE GeV data (see Discussion). The early X-ray data are also well described by our model, which satisfactorily connects with the second set of observations by \textit{Swift}.

The calculation of the spectral and lightcurve properties of the late afterglow proceeds following the adiabatic hydrodynamic evolution of the fireball. This point is particularly relevant because, once the overall parameters of Section~\ref{sec:spectra} and the time dependence of $\varepsilon_e$ and $\varepsilon_B$ of \Cref{tab:evolution_parameters} (as determined during the early phases) are set, our model predicts the spectral and lightcurve evolution of the late afterglow with no free parameters.

In the late phase at $t' > 10^4$~s, GeV-TeV spectral data are not available. Only optical and X-ray data can be considered, which in our scenario reflect the evolution of the synchrotron component. The X-ray flux evolution is well matched by our model also in this phase, predicting the observed temporal flux index up to $t' \sim 10^5$~s. 
Interestingly, after that, a slight continuous curvature appears in the calculated lightcurve due to the intervening spectral break of the synchrotron peak in the same energy band, which also supports the stability of the hard photon index reported  by \citet{Swift_williams_2023}. This interpretation excludes the presence of a flux steepening due to a jet break, and it is in agreement with the absence of a similar curvature in the late optical lightcurve. 
Furthermore, it is interesting to notice that - in this late phase at about $10^4 < t' < 10^5$~s - our model predicts a TeV flux compatible with the current and future gamma-ray observatories, which makes these long GRB events possibly detectable even after about 1 day from the initial triggers.

In the very late phase $t' > 10^5$~s, in our model, the exponential spectral break of the synchrotron emission at $\gamma_{\text{max}}$ continues
its transit through the X-ray band and predicts the softening of the integral flux intensity evolution. After $t' = 10^6$~s, the model starts to depart from the observed X-ray data in the latest phase of the GRB emission. The X-ray lightcurve at such late times may also be influenced by other processes, e.g., a late-time energy injection increasing $\gamma_{\text{max}}$ or a shallow spectral break rather than the exponential break applied here.
Indeed, this spectral effect affecting the lightcurve is not reported in the optical band, which is accurately predicted above $t' > 10^5$~s, in agreement with the data. However, when taking into consideration earlier data at $t' < 10^5$~s \citep[see][]{laskar_2023}, our model tends to overproduce such an optical emission. This situation is not uncommon in long GRB fireball modeling \citep[e.g.,][]{Racusin_energetics_2009}. The early optical emission may be influenced by a  reverse shock providing enhanced optical flux in the first phase, which then is softened and later modified by the transit of the characteristic frequencies \citep[e.g.,][]{sari_1999_optical_flash, Laskar2013_RS_GRB190427A, Zhang2023_reverse} or by the contribution of an emerging supernova.

In \Cref{tab:flux_time_evolution} we show a comparison between model and data of the temporal index $\beta$ assuming that the flux lightcurve behaves as $\propto t^{-\beta}$. The model shows a satisfactory agreement with the observed data.\\

\begin{table}
\centering
\caption{Comparison between model and data of the power-law flux evolution over time as $\propto t^{-\beta}$ of different energy bands in the lightcurve. Data are extracted from (a) \citet{LHAASO_2023}, (b) \citet{Tavani_2023}, (c) \citet{Swift_williams_2023}, (d) \citet{Fulton_optical_lightcurve}. In (e) the model curvature influences the estimation in this time interval.
}
\renewcommand{\arraystretch}{1.4}
\begin{tabular}{|cccc|}
\toprule
\setrow{\bfseries}Energy & $\mathbf{\Delta t'}$  &  \multicolumn{2}{c|}{\textbf{Index} $\mathbf{\beta}$} \\ 
 & [s]  &  Model & Data  \\ 
\hline
TeV & $20 - 200$  & $ 1.06 \pm  0.05$ & $1.115 \pm  0.012$$^a$\\
GeV  & $50 - 2\cdot10^4$ & $ 1.30 \pm  0.10$ & $1.300 \pm  0.200$$^b$ \\
\multirow{2}{*}{X-rays} & $1\cdot10^3 - 1\cdot10^4$ & $ 1.56 \pm  0.06$ & $1.498 \pm  0.004$$^c$\\
 & $1\cdot10^5 - 1\cdot10^6$ & $ 1.60 \pm  0.18$ & $1.672 \pm  0.008$$^{c,e}$\\
 Optical & $3\cdot10^5-2\cdot10^6$ & $ 1.57 \pm  0.03$ & $1.556 \pm 0.002$$^d$ \\
\toprule
\end{tabular}
\label{tab:flux_time_evolution}
\end{table}

\section{Discussion}
\label{sec:discussion} \label{sec:gammamax}

\noindent
In this work, our theoretical model has been computed by keeping constant the largest number of parameters during the GRB afterglow evolution in order to deduce the physical constraints.

The ISM density profile has been kept constant and homogeneous with $n_0 = 0.8\;\text{cm}^{-3}$ throughout all the phases of the GRB. 
This assumption is supported also by other works and previous analyses of long GRBs \citep[see e.g.,][]{PanaitescuKumar2002, oconnor_structured_jet}, but it may represent a simplification of the real configuration with an average density of the ISM during the early phases.

We also assume the presence of a cutoff at $\gamma_{\text{max}}$ in the distribution of accelerated particles, which suppresses the maximum energy of the synchrotron emission, affecting both the SEDs and the lightcurve at specific times. 
This physical feature can be directly investigated  with the AGILE data presented in \citet{Tavani_2023}.
We verified that at two specific time intervals  $t'=$ [47, 157]~s and $t'=$ [458, 608]~s, the AGILE GRID data moderately constrain the value of $\gamma_{\text{max}} \lesssim 4 \cdot 10^7$, which we adopt in our modeling as a constant parameter.
After these intervals, AGILE data do not constrain $\gamma_{\text{max}}$, which may increase in time.

In our analysis, the effect of internal $\gamma\gamma$ absorption is not detectable in the very early SEDs. Conversely, as shown in \Cref{fig:sed_AGILE_LHAASO_int1}, it marginally affects the early spectra at $\sim5$~TeV energies \citep[as seen also by][]{LHAASO_2023}.
This minor effect may be related to a physical detachment between the regions emitting high- and low-energy photon fields, possibly due to the specific spatial evolution of the outwardly propagating shock wave. \\

\subsection{Shock efficiencies $\varepsilon_e$ and $\varepsilon_B$}
\noindent
A crucial physical process in GRB afterglows is the conversion of the shock energy into nonthermal particle acceleration and magnetic field generation, characterized in this model by the two efficiencies $\varepsilon_e$ and $\varepsilon_B$. These two parameters simplify the mechanisms at the core of the afterglow emission, which can be very complex. The case of GRB~221009A is ideal to study this phenomenon, given the unusual burst duration and detailed spectral data available. Interestingly, we find that $\varepsilon_e$ and $\varepsilon_B$ cannot be constant in order to achieve a global fit of the experimental data. 
In our model, we adopt the following temporal power-law evolution for the entire GRB~221009A afterglow:
\begin{equation}
 \varepsilon_e \propto t^{0.19 \pm 0.02}  \quad \text{and} \quad  \varepsilon_B \propto t^{-0.84 \pm 0.04} \:, 
 \label{eq:3}
\end{equation}
which constitutes one of the major results of our analysis.
Our data fitting indicates that $\varepsilon_e$ slightly increases from 1.7\% to 5\% between a few seconds and $t' \simeq 5 \cdot 10^3$~s. Significantly higher values of $\varepsilon_e$ would shift the overall broadband emission to high energies, providing a disagreement with the data.
The quantity $\varepsilon_B$ evolves considerably over the same time intervals, decreasing from about $10^{-5}$ to $10^{-7}$, which is a range of values in agreement with the literature \citep[see discussion in][]{Duran2014_epsilonb, Santana2014_epsilonb, Nava2014}.
It is also interesting to notice that $\varepsilon_B \ll \varepsilon_e$ throughout the GRB afterglow evolution, which is also found in many GRBs with gamma-ray afterglows \citep[e.g.,][]{Beniamini2015}.
These time-dependent shock efficiencies successfully describe the spectral and flux intensity data of GRB~221009A up to late times, confirming an approach that has been explored in the past also for other long GRBs \citep[e.g. in][]{Ioka_2006, Granot_jets_2006, Maselli2014, Misra2021, laskar_2023}. However, their power-law time evolution may change at very late times with an increasing influence of other hydrodynamic and geometrical effects.

\vspace{5mm}

\subsection{Cooling and evolution of the spectral peaks}
\noindent
Our best modeling of GRB~221009A is based on the slow-cooling regime with $\gamma_{\text{min}} < \gamma_{c}$. We find that the predictions of the fast cooling are in contradiction with the afterglow data since the early times.

Additionally, it is interesting to note that during the early phases and throughout the afterglow, the synchrotron and inverse Compton peaks remain quite constant in time. We deduce this important feature from the GeV-TeV spectral data in intervals $T_1$-$T_3$, and subsequently from the X-ray and GeV data in interval $T_4$.
Focusing on the synchrotron cooling frequency $\nu_c$, we notice that the prediction for constant $\varepsilon_e$ and $\varepsilon_B$ implies $\nu_c \sim t^{-1/2}$, which is in contradiction with observations up to $t' \sim 10^4$~s. It is interesting to note that making $\nu_c$ weakly dependent on time is equivalent in our model to obtain that the combination
\begin{equation}
 \nu_c^{-1} \propto \Gamma  \: B^3  \: t^2 \propto \Gamma^{4}  \: \varepsilon_B^{3/2} \: t^2 
\end{equation}
remains nearly constant. 
Given the known dependencies of $\Gamma(t) \propto t^{-\frac{3}{8}}$ in the adiabatic slow-cooling scenario, either a more dissipative hydrodynamics $\Gamma(t)$ or a time evolution for $\varepsilon_B$ are required in order to preserve the quasi-constancy of the synchrotron and the inverse-Compton peak frequencies in the first times.
In our model, we have adopted a variable $\varepsilon_B$ according to Eq.(\ref{eq:3}), which implies critical frequencies evolving as shown in \Cref{fig:frequencies}.

\subsection{Energetics and jet breaks}
\noindent
Our initial isotropic-equivalent energy of the blast wave $E_{\text{iso,0}} \sim 7\cdot 10^{55}$ erg/s  is quite large, 
even though similar values have been reported by other authors \citep[e.g.,][]{2017AdAst2017E...5C, fermi_gbm_grb221009a, oconnor_structured_jet, LHAASO_2023, Ren_2023}.
The beaming-corrected isotropic-equivalent energy of the jet $E_{k}$ is given by 
$ E_{k} \simeq \frac{\theta^2}{2} E_{\text{iso,0}}$,
where the opening angle $\theta$ is often estimated by identifying the presence of 
achromatic jet breaks in the afterglow lightcurve \citep{Sari_jet_breaks, Granot_jets_2006}.
A first possible determination of a jet break has been proposed by \citet{LHAASO_2023}, reporting a steepening of the TeV gamma-ray lightcurve around $t' \simeq 670$~s. 
The deduced opening angle, $\theta \sim 0.6^{\circ}$, would be rather low compared to 
other GRBs \citep[e.g.,][]{frail_2001_jet_opening_angle, Atteia_2017_energetics_grb}.
In this case, the beaming-corrected shock energy of GRB~221009A, $E_{k} \sim 4\cdot 10^{51}$~erg, would be in agreement with the statistical distribution of long GRBs. 
However, the GeV data presented in this paper do not show any significant indications of a jet break up to $t' \sim 10^4$~s. Given the strong connection between GeV and TeV data, we deduce that a high-energy jet break at early times is unlikely.
Our model is also consistent with the data and does not predict an early jet break. As seen in Section~\ref{sec:lightcurve}, once the overall parameters in Section~\ref{sec:spectra} and the time dependence of $\varepsilon_e$ and $\varepsilon_B$ in \Cref{tab:evolution_parameters} are fixed, the spectral and lightcurve evolution are determined solely by the adiabatic hydrodynamic evolution of the fireball, even at very late times, within a consistent and unified scenario.

An alternative interpretation reported by \citet{davanzo_optical_jet_break_2022} suggests that the jet break may occur in the late afterglow phase at about $10^5$~s ($\sim 1$ day), when a steepening of the X-ray and optical fluxes occurs.
This implies a relatively large opening angle, $\theta > 15^{\circ}$, which would be larger than the common values in the literature \citep[supported also by][]{ixpe_grb221009a}. Consequently, the corrected isotropic-equivalent energy would be reduced\footnote{This interpretation is complicated by an actually nonsimultaneity of the jet break between optical and X-ray data, as indicated in \citet{Swift_williams_2023}.} 
to $E_{k} \sim 2\cdot 10^{54}$~erg.

Another interpretation by \citet{oconnor_structured_jet} interprets the X-ray and optical bending at $\sim$1 day as a geometrical effect due to the shallow energy profile of a  structured jet \citep[see also][]{sato2023, Zhang2023_rev_prot}. 
Such a scenario would reduce the model energy requirements, describing the late X-ray and optical afterglow after $\sim$0.8 day. The presence of a shallow structured jet may also provide modifications to the temporal dependence of the critical frequency \citep{Beniamini2020, Beniamini2022}. However, we do not discuss these effects in the current model as it would require further exploration of parameter correlations.

Given the nature of our precisely modeled data during the early phases of the GRB~221009A afterglow, we conclude that a canonical early jet break is supported neither by the data nor by this theoretical interpretation. A jet break might occur in the later phases or might not be canonically identified, as observed in other recent major events such as GRB~130427A \citep{depasquale_2016_GRB130427A}, GRB~190829A \citep{Dichiara2022_GRB190829A}, and GRB~190114C \citep{Misra2021, 2019Natur.575..455M}.


\section{Conclusions}
\noindent
The theoretical modeling of GRB~221009A requires an extraordinary approach. We modeled the early and late phases of the GRB~221009A afterglow within the context of the relativistic blast wave plus the synchrotron self-Compton scenario. The high-quality observational data, spanning from optical to X-rays and up to GeV-TeV gamma-ray energies, provide crucial constraints for the spectral and flux modeling of this very long GRB.

It turns out that a simultaneous fitting of all the spectral and flux data is not trivial and it is not achieved by any standard GRB model.
A physical model has to be confronted with a number of problematic features of the GRB~221009A afterglow, including: (a) the very precise spectral information first in the GeV-TeV range and later in the optical and X-ray bands that establishes the quasi-constancy of $\nu_c$ (rather than the standard behavior implying $\nu_c \sim t^{-1/2}$); (b) the overproduction of the GeV component in case of constant or relatively large $\varepsilon_B$; (c) the coherence between the X-ray, GeV and TeV lightcurves since the very early phases up to the late phases, indicating a highly constrained physical system evolving in a global way; (d) the bending of the X-ray lightcurve near $t' \sim 10^5$~s without invoking a jet break; (e) overall, the challenge of providing a consistent model explaining a large number of features at very different energies and timescales.

We provide such a model, investigating a comprehensive theoretical interpretation within the framework of a relativistic expanding fireball. 
Here, we emphasize its interesting properties: (1)  large values of the initial isotropic energy  $E_{\text{iso,0}}$ and of the initial bulk Lorentz factor $\Gamma_0$; (2) a homogeneous density profile with $n(r) \equiv n_0 = 0.8$~cm$^{-3}$; (3) the electron power-law index $p$ being constant; (4) a regime of adiabatic slow cooling throughout the entire afterglow; (5) a shock energy being progressively transferred to accelerated electrons with increasing efficiency (reflecting a very fundamental property of the physics of afterglow particle acceleration in this GRB); (6) a relatively small value of $\varepsilon_B$ varying from $10^{-5}$ to $10^{-7}$; (7) the relevance of a maximum energy of the electron distribution $\gamma_{\text{max}}$ that is constrained by GeV-TeV data during the early phases of the afterglow.

We also notice that the overall MeV-GeV-TeV datasets show the transition from a prompt-dominated phase to an afterglow-dominated phase up to $t' \sim 600$~s, with a rare clarity 
compared to other GRBs. Until this time, the activity of the inner engine contributing to the MeV emission overwhelms the radiation emitted by the optically thin region of the afterglow in the X-ray energy band. On the other hand, in the very early afterglow phase the physical high-energy component, being probably attenuated by internal $\gamma\gamma$ absorption with the prompt photon fields, it is then released and produce gamma rays as the opacity decreases. 
This is supported by the smoothness of the very early GeV-TeV lightcurve compared to simultaneous X-ray and MeV lightcurves, suggesting a different emission origin for these energy bands, the former representing an afterglow emission and the latter the turbulent activity of the inner engine leading to the prompt emission.

A key outcome of our analysis of the exceptional GRB~221009A is that, while the shock becomes progressively more efficient at energizing the non thermal population of radiating particles, the magnetic field energy density significantly decreases over time. This outcome in energy transfer and magnetic field generation may also be influenced by hydrodynamical effects acting on $\Gamma(r)$, beyond those considered in the model.
The time evolution of the microphysical parameters, supported by a precise fitting between model and multifrequency spectral and intensity data, is an important indication for future theoretical investigations of GRB~221009A and of other long GRBs.

\vspace{0.5cm}

\begin{acknowledgments}
\noindent
\textbf{Acknowledgments}\\
AGILE is a mission of the Italian Space Agency (ASI), with scientific and programmatic participation of INAF (Istituto Nazionale di Astrofisica) and INFN (Istituto Nazionale di Fisica Nucleare). 
This work was partially supported by the grant Addendum n.7 - Accordo ASI-INAF n. I/028/12/0 for the AGILE project. We are grateful to Marco Romani for his contributions on the subject of this paper that he presented in his Master's thesis. We thank an anonymous referee for stimulating comments on the manuscript.\\
\end{acknowledgments}


\appendix 
\section{Details on AGILE data}
In \Cref{tab:table_observ_windows} we provide the technical details concerning the analysis of AGILE GRID data.

\begin{table}[h!]
    \centering
    \caption{Technical details on the AGILE GRID lightcurve and spectra at GeV gamma-ray energies, shown in Figures~\ref{fig:lightcurve_only_data}-\ref{fig:sed_AGILE_LHAASO_early_times}. 
    On the left column ``Lightcurve'', we report information on each lightcurve bins, including time interval adopted for the integration window, signal detection, photon index (if statistics does not allow for the evaluation, we adopted 2.0), and energy flux. On the right column ``Spectra'', we report the overall integration times for the three simultaneous AGILE-LHAASO spectra adopted for the modeling and defined in \Cref{tab:evolution_parameters}. Notice that each spectrum (on the right side) corresponds to a specific time bin of the lightcurve (on the left side), except for interval $T_2$ that was covered by two spins of the AGILE satellite (lightcurve bins 2 and 3). Intervals $T_{-1}$, $T_0$ and $T_4$ are not shown as no AGILE GRID data were available. (a) indicates the photon index assumed for the computation of the lightcurve bin flux.}
    \renewcommand{\arraystretch}{1.2}
    \begin{tabular}{|ccccc|ccc|}
    \hline
        \multicolumn{8}{|c|}{\bf AGILE GRID} \\ 
        \hline
        \multicolumn{5}{|c|}{\multirow{1}{*}{\bf Lightcurve}} & \multicolumn{3}{c|}{\bf Spectra} \\ 
        ~ &&&&& GRID & LHAASO &\\ 
        \hline
        Bin \# & $\Delta{t'}$ & Detection & Photon index$^a$ & Energy flux  &$\Delta{t'}$  & $\Delta{t'}$ & Interval \\
         &  [s] & $\sigma$ & & [erg cm$^{-2}$ s$^{-1}$] & [s]  & [s]  &   \\ \hline
        1 & 47-100      & 42.3 & $1.83 \pm 0.07$ & $(4.4 \pm 0.4) \cdot 10^{-6}$ & 47-100 & 22-100 & \bf  T1  \\ \hline
        2 & 100 - 168   & 24.2 &  \multirow{2}{*}{$1.87 \pm 0.11$} & $(1.6 \pm 0.2) \cdot 10^{-6}$ & \multirow{2}{*}{100-608} &\multirow{2}{*}{100-674} & \multirow{2}{*}{\bf T2} \\ 
        3 & 458 - 608   & 15.8 & ~ & $(3.4 \pm 0.7) \cdot 10^{-7}$ & ~ && ~ \\  \hline
        4 & 903 - 1493  & 7.8 &  $1.87 \pm 0.44 $& $(7.4 \pm 2.2) \cdot 10^{-8}$ & 903-1493 & 674-1774 & \bf  T3  \\ \hline
        5 & 1788 - 5043  & 3.7 & $2.0$ & $(3.1 \pm 1.7) \cdot 10^{-8}$ &\multicolumn{2}{c}{} \\\cline{1-5}
        6 & 5047 - 16754  & 3.9 & $2.0$ & $(6.8 \pm 3.3) \cdot 10^{-9}$ &\multicolumn{2}{c}{}  \\   \cline{1-5}
    \end{tabular}
    \label{tab:table_observ_windows}
\end{table}

\newpage

\section{Characteristic frequencies of the model}

\begin{figure}[h!]
    \centering
    \includegraphics[width=0.5\textwidth]{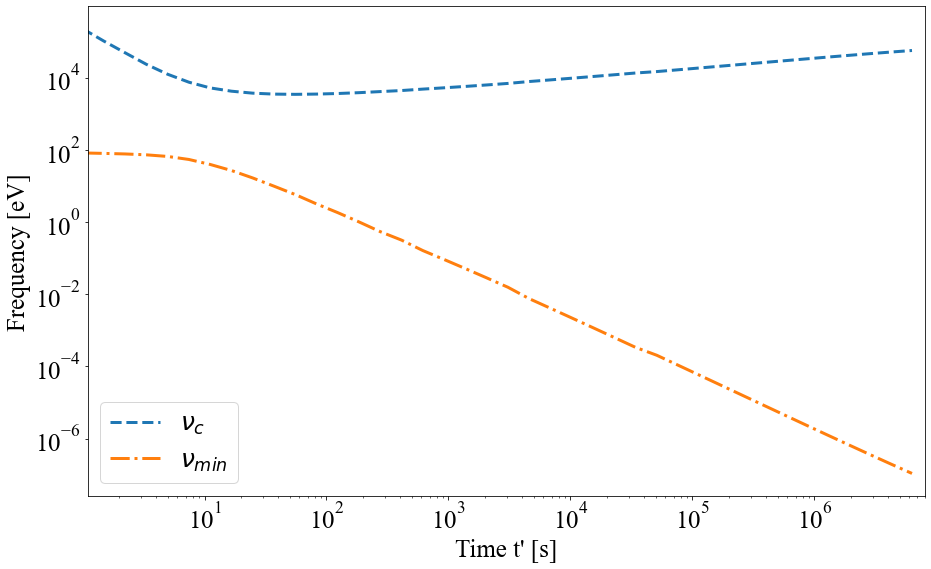}
    \caption{Time evolution of the calculated characteristic frequencies of the model of the GRB~221009A afterglow: the cooling frequency $\nu_c$ is shown with a dashed blue line, and the minimum frequency $\nu_{\text{min}}$ as a dashed-dotted orange line.}
    \label{fig:frequencies}
\end{figure}

In \Cref{fig:frequencies} we report the computed time evolution of the critical frequencies of the model.

\bibliography{biblio}{}
\bibliographystyle{aasjournal}

\end{document}